# Affordances and safe design of assistance wearable virtual environment of gesture

Didier FASS

*ICN Business School, Artem "augmented human" project*
*MOSEL LORIA Campus Scientifique BP239, 54506 Vandoeuvre-lès-Nancy, France*

**Abstract**

Safety and reliability are the main issues for designing assistance wearable virtual environment of technical gesture in aerospace, or health application domains. That needs the integration in the same isomorphic engineering framework of human requirements, systems requirements and the rationale of their relation to the natural and artifactual environment. To explore coupling integration and design functional organization of support technical gesture systems, firstly ecological psychologyprovides usa heuristicconcept: the affordance. On the other hand mathematical theory of integrative physiology provides us scientific concepts: the stabilizing auto-association principle and functional interaction. After demonstrating the epistemological consistence of these concepts, we define an isomorphic framework to describe and model human systems integration dedicated to human in-the-loop system engineering. We present an experimental approach of safe design of assistance wearable virtual environment of gesture based in laboratory and parabolic flights. On the results, we discuss the relevance of our conceptual approach and the applications to future assistance of gesture wearable systems engineering.
2015 Didier Fass.

## 1. Introduction

The safe design of virtual environments (VE) to aid in gestures, in particular the technical gestures is a challenge for the safety critical fields, such as aerospace or health. Whether simulators training or cockpit and systems maintenance, taking simultaneous account of 'human factors', technical factors and organizational factors is part of an overall concept of the "man in-the-loop" systems. VE, virtual reality and augmented reality, integrate humans into their processing by the generation of multimodal sensory-motor loops between actuators, sensors and computers. Performance and the overall security of these technological environments thus depend on the quality of the human systems integration (HIS) generated by the devices (more or less immersive and wearable) and these artificial sensorimotor loops in operational.

### 1.1. Reclaiming virtual environment: from ecology to integrative biology

The main difference between virtual environment systems and conventional vis-à-vis computer interactive systems is the fact that human as is a part of virtual environment – so called"participant" [1]. Participant means that the human is integrated into the operation by the generation of multimodal sensorimotor loops between actuators, sensors and computers. Thus wearable virtual environment can be seen simultaneously as an extension of the body and an increase of the environment that enhance human capabilities and performances.

Traditional approaches rely on analytic and a reductionist method. They propose an abstraction of the human based on a mechanical or computational model.An important aspect toconsider inthe design ofwearableassistance system is that the body's relationship with natural or artificial environment is structurally discrete and functionally continuous.

A shift indeed is need from classical epistemological interaction theory that grounds interaction designs, based on human machine reductionism and computing, to a theory ground on integration and related concepts, based on whole and a non-reductionist epistemology. Since we know that a biological system and a technical system are different by nature and organization, that requires to think human machine systems complexity inside a theoretical and practical framework which explains the functioning of a given system by those of its structural components coupling and their dynamical organization, fitted to a given context. The groundingdesign framework must ensure the ecological and economic consistency of the engineered system with the integrated human "sensorimotor and cognitive" body. This requiresrethinking thevirtualenvironment in thefieldsof ecology andintegrative biology.

*1.2. Concepts: affordance, stabilizing auto-association and functional interaction*

Performance and overall security depend on the quality of human-system integration (HSI) generated by these artificial sensorimotor loops. For ensuring a reliable and safeintegration of the human and the artificial systemwe face to key issues:
- Definingprinciple of ecological virtual environment design;
- Grounding human system integration engineering and the biocompatibility and biointegration of the artefacts principles on the biological nature of human.

To set the principle of ecological virtual environment designthe concept of affordance invented by James Gibson [2] perception psychologist is heuristic.Gibson's affordance means operable properties between the world and an individual (person or animal). For Gibson in its ecological approach, the affordance is: " a combination of physical properties of the environment that uniquely suited to a given animal-to his nutritive system or his action system or his locomotor system […] This is nothing to reemphasize that perception of the environment is inseparable from of one's own body- that egoreception and exteroreception are reciprocal."

This concept reflects the ability of one or more components of the environment to generate an individual couplings adapted to its environment. Specifically, it refers to the actionable properties between the world and the individual in non symetrical flow.That refers to the ability of one or more components of the environment natural or artificial to generate suitable couplings - ecological and physiological-, consistent of an individual with this environment.

Note that classical heuristical [3] use of the affordance concept in design refers to a symmetrical interaction between the artefacts and the individual.

This concept of is consistent with the principles of the Mathematical Theory of Integrative Physiology (MTIP) set by Gilbert Chauvet [4]. Chauvet defined and validated formally two main theoretical principles for modelling a hierarchical representation of the biological systems functional organization [5]:
- Stabilizing auto-association principle (PAAS): for the biological system, the PAAS represents a general principle of organization, which describes why two structures tend to associate themselves, carrying out a new function. In biological systems associated units exist because their functional process associated to another one become more stable.
- Functional interaction: MTIP introduces the principles of a functional hierarchy based on structural organization within space scales, functional organization within time scales and structural units that are the anatomical elements in the physical space. It faces the problem of structural discontinuity by introducing functional interaction from structure-source into structure-sink, as a coupling between the physiological functions supported by these structures. The area of stability of the system - structure-source structure-sink and the functional interaction - is larger than the areas of stability of structure-source and structure-sink considered separately. In other words, the increase in complexity of the integrated system corresponds to an increase in stability [5].

These concepts and principles ground our conception of the human systems integration and raised capacity and human performance by artificial coupling or affordances loops. These are the organizational and hierarchical integration principles underpinning our experimental approach.

*1.3. Gesture and assistance of gesture*

Gesture is a goal-oriented movement. It is an action that result from a sensorimotor and cognitive high level of integration with the surrounding environment.

The behavior of the human operator is the result of a set of physiological functions that can be classified into three broad categories: perception, decision and action. To these threebasic functions, addscontrol (control loop), which aims toensure thatthe action isreliablyin thefield of securityof the overall system, for both the human operator and the operated system.

For stableactionsituation thatlead to thegood gestureandthe desiredoperational performance, make surethat the operatorperceives, decides and controlscorrectly. Sensorimotor patterns provided by the multimodal environment must be adapted to human capabilities and properties. The correct design of assistance werable virtual environmentsof gestures requires to take into account both quality of the sensorimotor patterns generated by devices and physical properties of these wearable devices.

To describe and model these elements of construction, we linkwith the concepts of Gibson affordance and Chauvet PAAS and functional interaction. We hypothesize the generation of combinations of biocompatible physical affordant properties in operational situation by the werable support system. That is to say adapted to perceptual-motor system of the individual, leading to a set of hand movement for which a stable situation is obtained.Stabilityheremeans that thegestureisrealizedwithin the integrated domain ofperformance and reliabilityof the prescribed action. This actionresult froma set of functionalcouplingsbased ona specific integrative organization of both biological and technical structural elements, shapes or forms, anddynamics.

The definitionof the support systemconsists in findingtechnical solutionsbythe means to maintainthis human system integration: the design of wearble devices and artificial sensori-motor patterns. These solutions cantherefore take account ofperceptualabilities, cognitiveand motor skills of the human bodily integrated tothe virtual environmentand coupled to the space of action by "affordances" and/or functional interactions.

## 2. Human systems integration and wearable virtual environment engineering

What structural and functional elements of virtual environments generate sensorimotor adapted patterns?

What organizationof these elements,which design, ensuring human systems integration?Performance?Comfort?Safety? Overall reliability ?

To answer these two questions we propose:
- Looking for classes of affordant elements of structure, of beahvior and of evolution that based the design of these dedicated wearable environments and demonstrate their ability to generate functional emergence of reliable technical gesture;
- Searchingdynamicorganizational principles ofwearablehumanin-the-loop systems and of human systems integration relatedhumanmotor performance.

*2.1. Design and system engineering*

The research of the "affording" elements and their theoretical and methodological dynamic functional organizational principles is a major challenge for the integrative design and engineering of systems of assistance to gestures. They determine motor performance by the generation of suited affordances, grounding of HSI. These elements are multimodal structural and functional elements of organization whose potentials of affordance are involved in perceptual-motor couplings of the individual and the artefacts.

The design of these gestures support systems should meet two requirements complementary and not reducible one to the other: first, search for classes of "affording" elements of structure and shape in the design of these dedicated VE and demonstrate their ability to generate coupling by functional interactions; secondly, search the principles of dynamic organization of the systems architecture to support actions that ensure the human systems integration and motor performance.

*2.2. Human system integration domain*

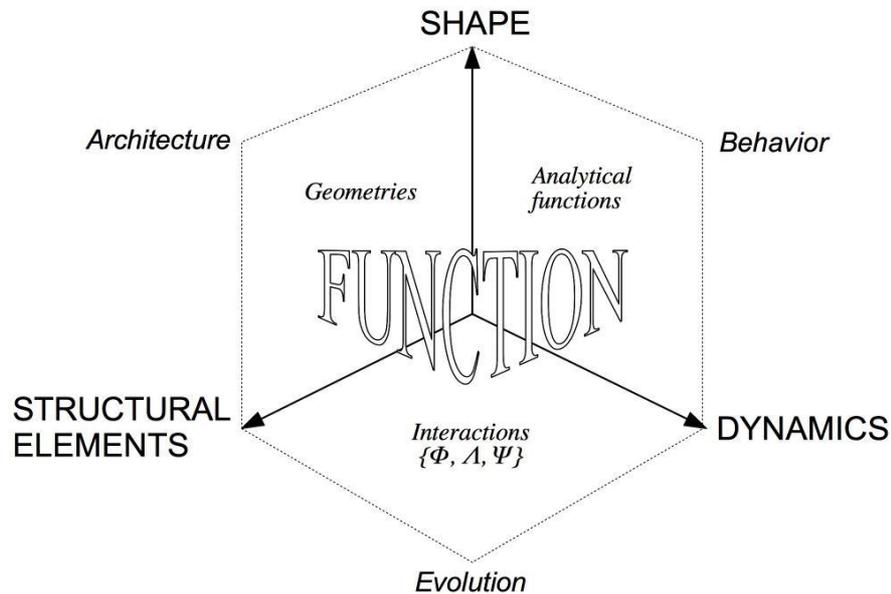

Fig. 1. Our overall system design general conceptual isomorphic framework: System function results from the integrative organization of different structural elements, shapes and dynamics according there space and time scales relativity and specific qualitative and quantitative measurement units.

Since technical systems are mathematically grounded and based on physical principles, human in-the-loop systems (HITLS) needs to be considered in mathematical terms. There are several necessities to make HIS and augmented human reliable [6].

Necessity 1 – Designing a human in-the-loop systems is to couple two systems from different domains organized and grounded on different principles theory and framework: biological, physical, numerical.

Necessity 2 – human in-the-loop systems design is a global and integrative model based method ground on Chauvet's Mathematical Theory of Integrative Physiology and domain system engineering.

Necessity 3 – Modelling assistance systems of gesture and human system integration is to organize the required hierarchically structural elements, shapes functions and their interactional dynamics according an architectural principles, behavioural needs of performance and efficiency and evolutionary needs.

Consequently, designing an assistance wearable virtual environment of gesture following human system integration is to organize hierarchically and dynamically human and artefat coupling. This requires a new domain engineering approach for requirements and specification based on biological user's needs and functions.

For modeling integrated human machine systems, we have grounded our new epistemic framework on isomorphism (Fig. 1). We have found new general categories of system elements or affordances that allow the same representation framework for integrating two systems different by nature. Our isomorphic epistemic framework is composed by three main isomorphic categories of interlocked elements: structural element, shape and dynamic.

As previously we were able to define three pair of isomorphic relation: Architecture{structural elements, shape}, Behavior{shape, dynamics} and Evolution{structural elements, dynamics}. Thus architecture is describable by a set of possible geometries, from Euclidian to no-Euclidian and other, behavior is describable by a set of functional analysis and algorithm, and evolution is describable by a set of modal interactions where each interaction might be compose on three modal parameters: physical, logical and physiological or behavioral dimensions according to Chauvet's theoretical integrative physiology.

*2.3. Virtual environment engineering*

In addition, there are several main requirements categories to make human systems integration and wearable virtual environment design safe and efficient. They address the technology - virtual environment-, sensorimotor integration and coherency.

**Requirement 1: Virtual environment is an artifactual knowledge based environment**
As an environment, which is partially or totally based on computer-generated sensory inputs, a virtual environment is an artificial multimodal knowledge-based environment. Virtual reality and augmented reality, which are the most well known technologies of virtual environments, are obviously the tools for the augmented human design and the development of human in-the-loop systems. Knowledge is gathered from interactions and dynamics of the individual-environment complex. It is an evolutionary, adaptive and integrative physiological process, which is fundamentally linked to the physiological functions with respect to emotions, memory, perception and action. Thus, designing an artifactual or a virtual environment, a sensorimotor knowledge based environment, consists of making biological individual and artifactual physical system consistent. This requires a neurophysiological approach, both for knowledge modelling and human in-the-loop design.

**Requirement 2: Sensorimotor integration and motor control ground behaviour and skills**
Humans use multimodal sensorimotor stimuli and synergies for interacting with their environment, either natural or artificial (vision, vestibular stimulus, proprioception, hearing, touch, taste…) [7]. When an individual is in a situation of immersive interaction, wearing head-mounted display and looking at a three-dimensional computer-generated environment, his or her sensorial system is submitted to an unusual pattern of stimuli. This dynamical pattern may largely influence the balance, the posture control (Malnoy& al. 1998), the spatial cognition and the spatial motor control of the individual. Moreover, the coherence between artificial stimulation and natural perceptual input is essential for the perception of the space and the action within. Only when artificial interaction affords physiological processes is coherence achieved.

**Requirement 3: Coherence and HIS insure the human-artefact system performance, efficiency and domain of stability**
If this coherence is absent, perceptual and motor disturbances appear, as well as illusions, vection or vagal reflex. These illusions are solutions built by the brain in response to the inconsistency between outer sensorial stimuli and physiological processes. Therefore, the cognitive and sensorimotor abilities of the person may be disturbed if the design of the virtual environment does not take into account the constraints imposed by human sensory and motor integrative physiology. The complexity of physiological phenomena arises from the fact that, unlike ordinary physiological systems, the functioning of a biological system depends on the coordinated action of each of the constitutive elements (Chauvet 2002). This is why the designing of a virtual environment as an augmented biotic system, calls for an integrative approach.

Integrative design strictly assumes that each function is a part of a continuum of integrated hierarchical levels of structural organization and functional organization as described above within MTIP. Thus, the geometrical organization of the virtual environment structure, the physical structure of interfaces and the generated patterns of artificial stimulations, condition the dynamics of hierarchical and functional integration. Functional interactions, which are products or signals emanating from a structural unit acting at a distance on another structural unit, are the fundamental elements of this dynamic

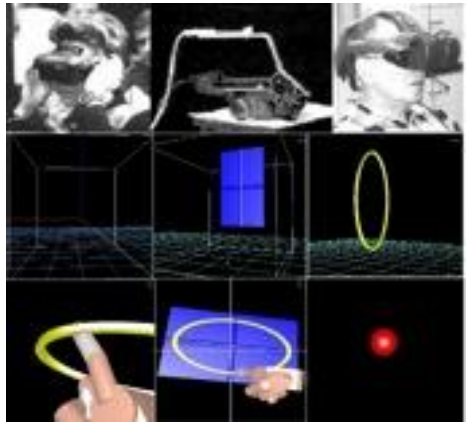

Fig. 2. Examples of different structural and functional affordances for virtual environment design.

## 3. Experience

To highlight these "affording elements" and their dynamic organization for assistance to gestures, we have defined *a priori* main classes of elements that constitute the virtual environment (Fig. 2.):
- Virtual reality;
- Augmented reality (see-through);
- type of head mounted display (immersive or not, optical resolution -high or low, field of view -wide or narrow, light or heavy, well balanced or not);
- Visual structure of the artificial space (allocentric or egocentric, background colour),
- Shape and plan of the movement; visual feedback of movement (anthropomorphic - hand - or abstract - ball),
- Dynamic quality of this visual feedback (remnant, persist);
- Sensorimotor coupling (visuo-proprio-kinesthetic or visuo-haptic);
- Spatial orientation of the gesture in the three dimensional space (frontal, sagittal and horizontal).

Taking into account this classification, we have established an experimental protocol based on behavioral analysis of the graphical gesture - drawing of ellipses in bi-dimensional or three-dimensional space. This Protocol follows an incremental complexity of dynamic organization allowing an intuitive and non-verbal learning task and the use of our gesture support system. Ten volunteers conducted at the laboratory (Fig. 3) of the drawings of ellipses in 3D space and on 2D support, in situation control bareheaded then wearing aid system to gesture in virtual reality and augmented reality for each of the 289 steps of complexity.

Three of the subjects also performed parts of this protocol in hypergravity and microgravity during a French National Space Centre (CNES) parabolic flight mission.

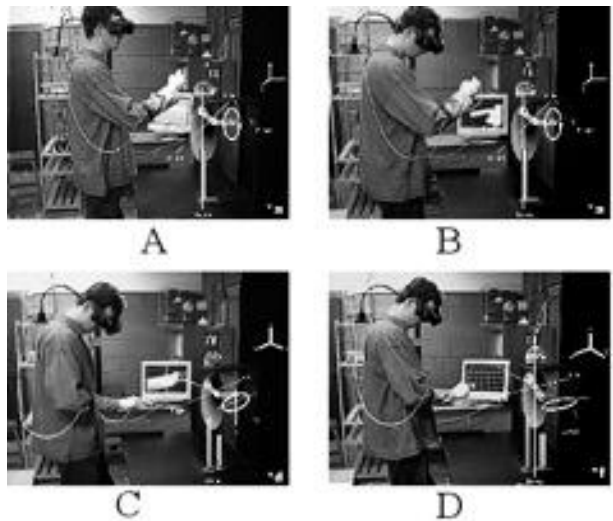

Fig. 3. Graphical gesture of ellipse drawing in the 3D space is performed and analysed in different configurations, more or less complex, of immersive virtual environment assisted drawing ellipses: A- SV ellipses and neutral and coloured background, B- SV ellipses and anthropomorphic visual feedback of movement (artificial hand), C- TF and model of ellipse insert in its plan of movement without visual feedback of movement, D- TH ellipses and abstract representation visual feedback of movement (ball).

Data processing allowed us to calculate four categories of variables characterizing the graphic gesture: cinematic, position, orientation and form. The exploratory multivariate statistical analysis (Princpal Component Analysis and hierarchical classification) confirms the existence of classes of affording elements and shows the more or less significant relative influence on the human systems integration, the performance and the quality of the gesture. Return of movement, visual or visuo-haptic, is the essential element for the help to the gesture in virtual environment. Spatial orientation of the movement mostly affects the accuracy of orientation of the gesture. It shows the influence (by postural and vestibular functional interactions) of the design of helmets, even off, on the gesture.Moreover gesture itself is an excellent marker of the design of virtual environment elements and of the human systems integration assessment.(For detailed protocoland resultssee [6])

## 4. Conclusion

In agreement with the Chauvet's theoretical biology principles, PASS and functional interaction, an artificial structural or functional element of the system of gesture support is an affordance when it generates a functional or physiological interaction strengthening the stability of sensorimotor coupling in improving motor performance of the operator and global behavior of the human machine system. These results areconsistent with the ecological conceptof affordance define by Gibson. The development oftechnicalgesturessupport systemsmust be based onthese principlesexperimentally validated.

Based on that theoretical framework and experimental results, our future workaddresses theepistemologicaland formalvalidationof the general frameworkfor modelinghumansystems integration. Our challenge is to develop correctness by construction system engineering for assistance wearable virtual environment of technical gesture.

This isa reliabilityissues forapplications in aerospaceand medicalsafety critical human in-the-loop systems

**Acknowledgements**

Research supported by Communautéurbaine du grand Nancy (CUGN)